\documentclass[lettersize,journal]{IEEEtran}
\usepackage{amsmath,amsfonts}
\usepackage{algorithmic}
\usepackage{algorithm}
\usepackage{array}
\usepackage[caption=false,font=normalsize,labelfont=sf,textfont=sf]{subfig}
\usepackage{textcomp}
\usepackage{stfloats}
\usepackage{url}
\usepackage{verbatim}
\usepackage{graphicx}
\usepackage{cite}
\usepackage{color}
\usepackage{units}
\begin{document}

\title{Recent Trends and Future Prospects of Neural Recording Circuits and Systems: A Tutorial Brief}

\author{Jinbo~Chen,~\IEEEmembership{Graduate~Student~Member,~IEEE,}
	Mahdi~Tarkhan,~\IEEEmembership{Member,~IEEE,}
	Hui~Wu,~\IEEEmembership{Graduate~Student~Member,~IEEE,}
	Fereidoon~Hashemi~Noshahr,
	Jie~Yang,~\IEEEmembership{Member,~IEEE,}
	and~Mohamad~Sawan,~\IEEEmembership{Fellow,~IEEE}
	\vspace{-0.6cm}

\thanks{Manuscript received 2022; revised 2022. (Corresponding author: Jie Yang, Mohamad Sawan.)}
\thanks{Jinbo Chen is with the Zhejiang University, Hangzhou, Zhejiang, China and with the Cutting-Edge Net of Biomedical Research And INnovation (CenBRAIN), School of Engineering, Westlake University, Hangzhou, Zhejiang, China.}
\thanks{Mahdi Tarkhan, Hui Wu, Jie Yang, and Mohamad Sawan are with the Cutting-Edge Net of Biomedical Research And INnovation (CenBRAIN), School of Engineering, Westlake University, Hangzhou, Zhejiang, China (e-mail: yangjie@westlake.edu.cn, sawan@westlake.edu.cn).} 
\thanks{Fereidoon Hashemi Noshahr is with the Microchip Technology Inc., Canada.} 
}



\maketitle

\begin{abstract}
Recent years have seen fast advances in neural recording circuits and systems as they offer a promising way to investigate real-time brain monitoring and the closed-loop modulation of psychological disorders and neurodegenerative diseases. In this context, this tutorial brief presents a concise overview of concepts and design methodologies of neural recording, highlighting neural signal characteristics, system-level specifications and architectures, circuit-level implementation, and noise reduction techniques. Future trends and challenges of neural recording are finally discussed.
\end{abstract}

\vspace{-0.4cm}

\section{Introduction}
\begin{figure}[t]
	\centering
 	\includegraphics[width=3.5in]{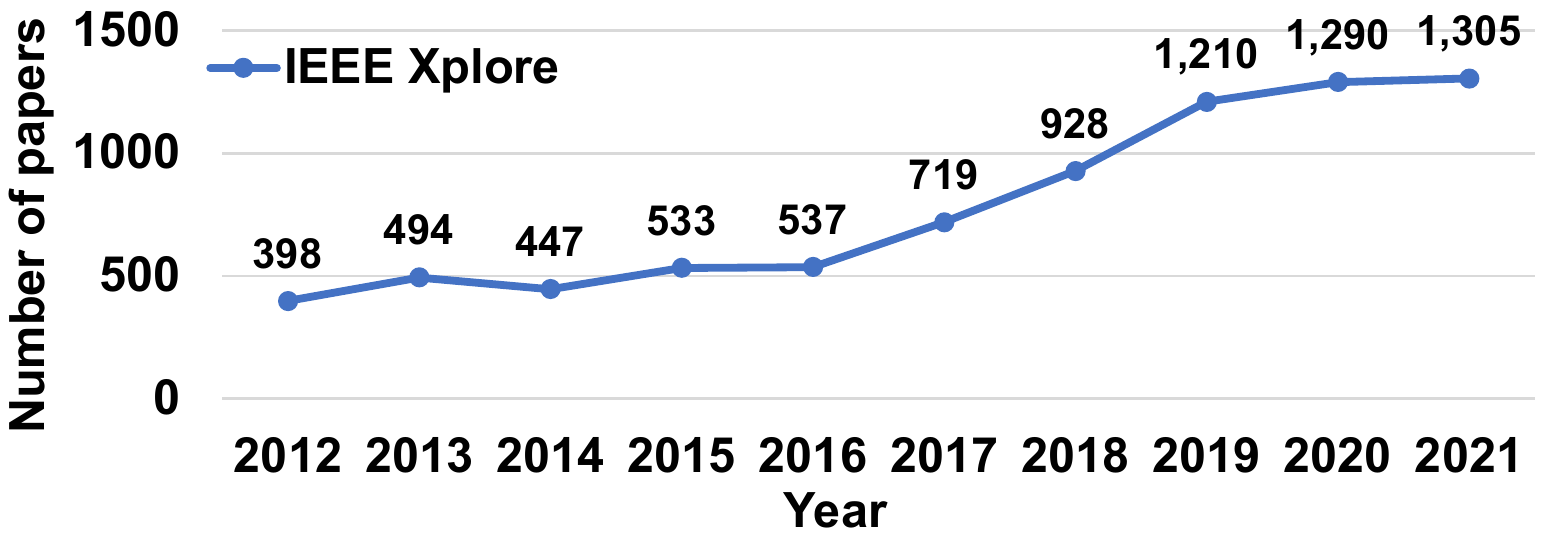}
	\vspace{-0.8cm}
	\caption{Neural recording related papers published in IEEE Xplore over the last 10 years (using keywords “neural recording”).}
	\label{papersXplore}
\end{figure}

\IEEEPARstart{N}{eural} recording circuits and systems have undergone rapid development in recent years as they are indispensable to brain–machine interface (BMI) systems that are critical for the brain science discovery and treatment of psychological disorders and neurodegenerative diseases \cite{RN652, RN650}. Fig. \ref{papersXplore} shows the rapidly increasing number of neural recording related papers published on IEEE Xplore over the last 10 years. Signals resulting from activities of brain neurons, such as electroencephalogram (EEG), electrocorticography (ECoG), local field potential (LFP), and action potential (AP), are recorded by dedicated neural recording analog front-end (AFE) with penetrating or surface electrode arrays. Table I presents characteristics of various neural signals. 

To record neural signals with relatively small amplitude and low-frequency, neural recording systems must fulfill certain specifications. Firstly, as there are approximately $10^{11}$ neurons in the human brain \cite{RN647}, multi-channel neural recording is required to achieve high spatial and temporal resolution. Therefore, the area per channel has to be sufficiently small. Limited battery capacity and wireless power transfer efficiency of recording systems make power consumption another crucial restriction. Moreover, low on-chip power density is required to solve the tissue heating issue. Thirdly, input-referred noise from the inherent noise of tissue, electrodes, and circuits is expected to be small enough to avoid overwhelming small amplitude neural signals. Remaining specifications include high electrode offset tolerance to avoid saturation of neural recording amplifiers, high common-mode rejection ratio (CMRR) to eliminate common-mode interference, and high input impedance to minimize amplifier gain attenuation. Additionally, fast saturation recovery is necessary to prevent the impact of stimulation transients and motion artifacts. Table II summarizes the target specifications of invasive and non-invasive neural recording systems \cite{RN650, RN6}.

The remaining sections of this tutorial brief are arranged as follows. Section II presents neural recording system-level architecture, followed by circuit-level implementation explained in Section III. Noise reduction techniques are described in Section IV. Section V investigates future trends and challenges of neural recording. Finally, Section VI concludes the tutorial brief.
\vspace{-0.5cm}

\begin{table}[t]
	\vspace{-0.3cm}
	\label{table_1}
	\caption{Characteristics of neural signal.}
	\vspace{-0.3cm}
	\resizebox{\linewidth}{!}{%
		\begin{tabular}{|c|c|c|c|} 
			\hline
			Signal Type      & Recording Type & Amplitude                                                 & Frequency                                                  \\ 
			\hline
			Extracellular AP & Invasive       & \begin{tabular}[c]{@{}c@{}}0.05 - 1 mVpp\\\end{tabular}   & \begin{tabular}[c]{@{}c@{}}100 Hz - 10 kHz\\\end{tabular}  \\ 
			\hline
			Intracellular AP & Invasive       & \begin{tabular}[c]{@{}c@{}}10 - 70 mVpp\\\end{tabular}    & \begin{tabular}[c]{@{}c@{}}100 Hz - 10 kHz\\\end{tabular}  \\ 
			\hline
			LFP              & Invasive       & \begin{tabular}[c]{@{}c@{}}0.5 - 5 mVpp\\\end{tabular}    & \begin{tabular}[c]{@{}c@{}}1 mHz - 500 Hz\\\end{tabular}   \\ 
			\hline
			ECoG             & Invasive       & \begin{tabular}[c]{@{}c@{}}0.01 - 1 mVpp\\\end{tabular}   & \begin{tabular}[c]{@{}c@{}}1 mHz - 200 Hz\\\end{tabular}   \\ 
			\hline
			EEG              & Non-invasive   & \begin{tabular}[c]{@{}c@{}}0.01 - 0.5 mVpp\\\end{tabular} & \begin{tabular}[c]{@{}c@{}}1 mHz - 200 Hz\\\end{tabular}   \\
			\hline
		\end{tabular}
	}
	\vspace{-0.3cm}
\end{table}

\begin{table}[t!]
	\centering
	\label{table_2}
	\caption{Target specifications of neural recording systems.}
	\vspace{-0.3cm}
		\begin{tabular}{|c|c|} 
			\hline
			Parameters                           & Target Specifications                                                                      \\ 
			\hline
			Area (per channel)                   & \textless{}~0.05~$mm^{2}$                                                                    \\ 
			\hline
			Power (per channel)                  & \textless{}~50 $\mu$W                                                                       \\ 
			\hline
			Bandwidth                            & Related with signal type                                                                   \\ 
			\hline
			Input-referred noise (per
			channel) &  \textless{} 10$\mu$$V_{rms}$                                                                  \\ 
			\hline
			Electrode offset tolerance           & \textgreater  ~$\left | \pm 50 \right |$~mV  \\ 
			\hline
			CMRR                                 & \textgreater~80 dB                                                                                     \\ 
			\hline
			Input impedance                      &  \textgreater~100 M$\Omega$                                                            \\
			\hline
		\end{tabular}
	\vspace{-0.6cm}
\end{table}

\section{Neural Recording System Architecture}
\subsection{Time-Division Multiplexing Based System Architecture}
\begin{figure}[t]
	\centering
	\includegraphics[width=2.3in]{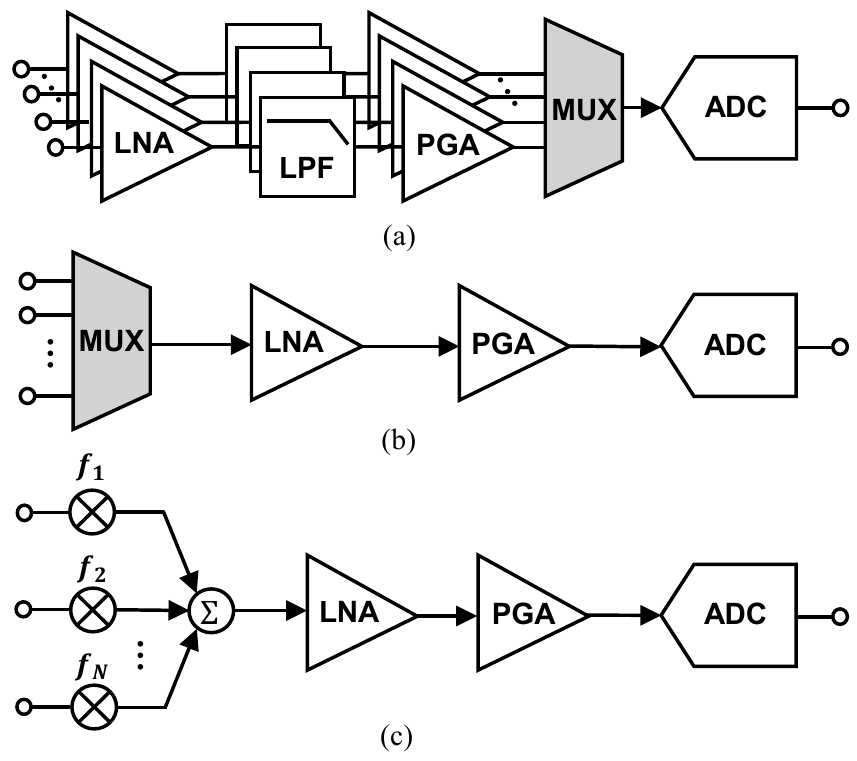}
	\vspace{-0.3cm}
	\caption{Neural recording system multiplexing architecture: (a) Time-division, (b) rapid time-division, (c) frequency-division.}
	\label{neuralsystem}
	\vspace{-0.7cm}
\end{figure}

Traditional multi-channel neural recording systems include low-noise amplifiers (LNA), multiplexers (Mux), and analog-to-digital converters (ADC). The LNA amplifies neural signals with narrow bandwidth and low amplitude. The ADC digitalizes amplified analog signals. Mux selects multi-channel input signals and transmits them to one AFE output line. The widely used architecture shown in Fig.~\ref{neuralsystem}(a) employs a single ADC for all channels. An LNA is used in each channel to amplify the signals. The multi-channel signals are transmitted to the ADC through an analog multiplexer using time-division multiplexing (TDM). In \cite{RN48}, an example of this design is described. However, when the number of channels rises, the ADC sample rate also increases, resulting in greater power consumption. Furthermore, multi-channel analog signals are susceptible to crosstalk noise in analog multiplexers, necessitating additional design strategies to enhance noise margin. The chip area per channel is still too large to integrate thousands of channels and beyond.

To realize high-density recording, the notion of rapid multiplexing is introduced to investigate TDM to swiftly record numerous electrode sites utilizing a single AFE circuit without preamplification \cite{RN356}. The recording architecture based on rapid multiplexing at electrodes shown in Fig.~\ref{neuralsystem}(b) could significantly reduce the number of amplifiers and ADCs. Similar recording architectures are also explored in \cite{RN601, RN602}. This multiplexing technique can also be co-designed with direct digitalization converters explained in the following sections. However, the effectiveness of this multiplexing strategy is dependent on electrode characteristics. Therefore, how to minimize electrode input offset and noise aliasing at the electrode-tissue interface becomes critical challenges.
\vspace{-0.5cm}

\subsection{Frequency-Division Multiplexing Based System Architecture}
Due to the performance limitation of the ADC and the necessity for a broad dynamic range to capture neural signals with considerable amplitude fluctuation across channels, TDM-based AFE consumes high energy and takes up large on-chip area. The frequency-division multiplexing (FDM) method has been used in neural recording to develop high-density and low-power AFEs to address these challenges.

The architecture of an FDM-based system is shown in Fig.~\ref{neuralsystem}(c). The frequency-modulated signal of each channel is converted and modulated to a specific frequency value. The cross-channel frequency-modulated signal outputs are aggregated into a signal wire and sent to a single amplifier and an ADC for signal amplification, demodulation, and processing. This permits the reduction of redundant cable wires and circuits, resulting in energy-efficient small area systems. Furthermore, frequency modulation (FM) ExG signals can also filter motion and electromagnetic interference (EMI). Recently, a distributed multi-channel FM-based AFE and FM-ADC architecture for ExG signal recording have been proposed in \cite{RN64}. Another work \cite{RN493} has also presented a technique that utilizes a single shared ADC among recording channels exploiting FDM with Walsh-Hadamard-based orthogonal sampling.
\vspace{-0.3cm} 

\section{Neural Recording Circuit Implementation}
\subsection{Neural Recording Amplifiers}
Small amplitude and low-frequency neural signals need to be amplified before digitalized by an ADC. Generally, several factors should be considered in neural amplifier designs, including adequate gain, suitable bandwidth, sufficient signal-to-noise ratio (SNR), high CMRR and input impedance, minimized energy consumption, and compact chip area \cite{RN7, RN637, RN635}. The two most prevalent topologies are AC-coupled and DC-coupled neural amplifiers. 

\begin{figure}[t]
	\centering
	\includegraphics[width=3.5in]{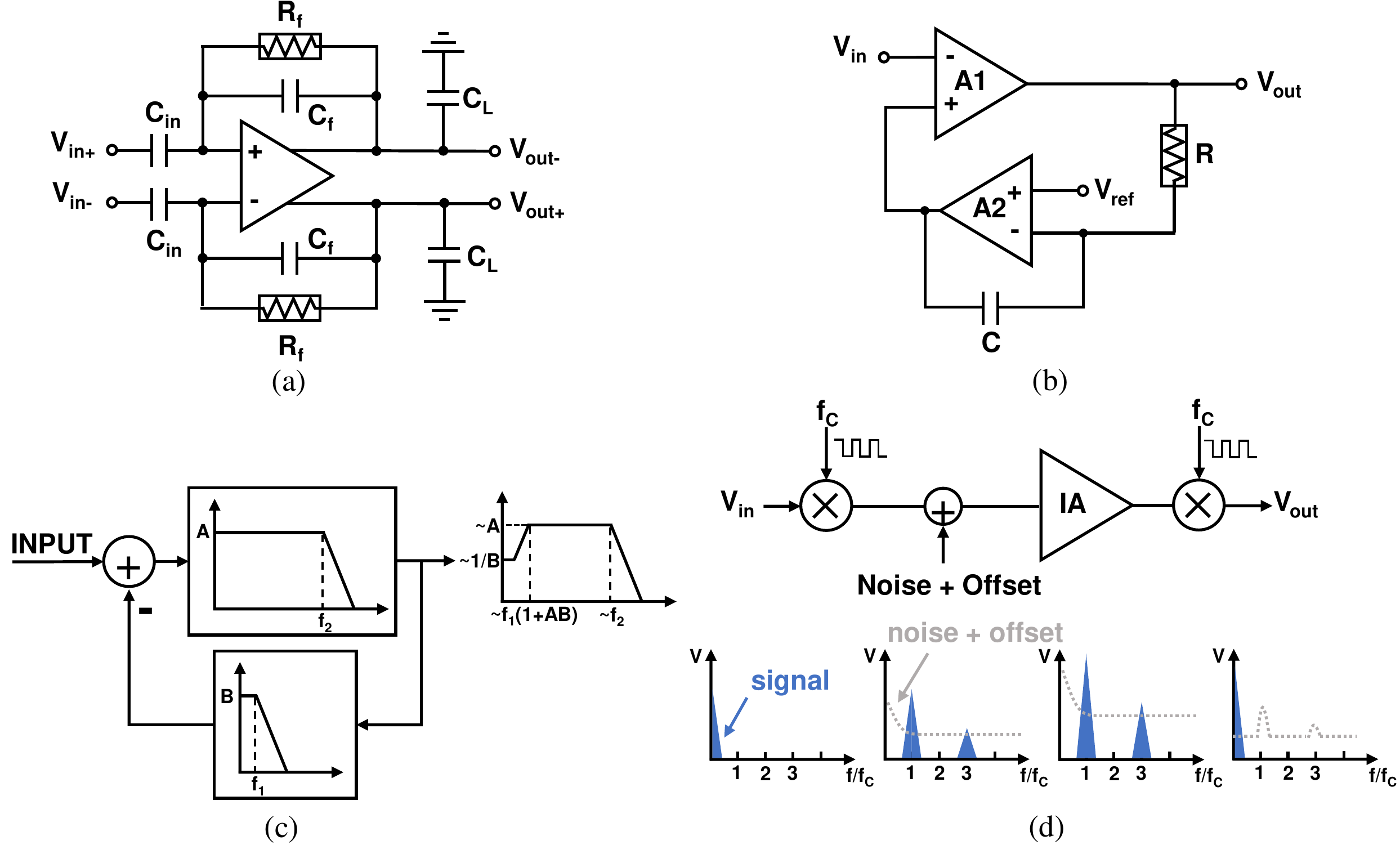}
	\vspace{-0.7cm}
	\caption{Neural amplifier architectures: (a) AC-coupled, (b) DC-coupled, (c) the idea of utilizing a low-pass filter in the feedback loop of DC-coupled amplifiers to build a high-pass filter, (d) diagram of chopping technique to alleviate 1/f noise and offset \cite{RN635}.}
	\label{neural_amplifier}
	\vspace{-0.7cm}
\end{figure}

\subsubsection{AC-Coupled Neural Amplifiers}
The capacitively coupled inverting amplifier shown in Fig.~\ref{neural_amplifier}(a) is a typical AC-coupled neural amplifier architecture \cite{RN633}. This amplifier's gain is equal to $\nicefrac{C_{in}}{C_{f}}$. A considerable capacitor $C_{in}$ at the amplifier's input guarantees high gain and blocks input DC offset, while diminishing the input impedance. Pseudo-resistors connected in parallel with feedback capacitors are employed to produce a high-pass pole with a small area. Furthermore, the input transistors are usually in big size to minimize flicker noise. This topology's major disadvantage is its considerable area overhead. To ensure a low-frequency high-pass pole, the feedback capacitor $C_{f}$ must be large, making the input capacitor $C_{in}$ unable to shrink significantly. For example, input capacitors in the range of tens pF are required to obtain a high-pass pole around 1 Hz with a gain of 40 dB. These large capacitors occupy significant on-chip area and lower the input impedance. Using two or three gain stages is an alternative approach to minimize input capacitors' size and reduce the chip area. However, this approach incurs extra power consumption.

\subsubsection{DC-Coupled Neural Amplifiers}
The main issue of DC-coupled neural amplifiers is eliminating input DC offset. Fig.~\ref{neural_amplifier}(c) illustrates an implementation that incorporates a low-pass filter in the feedback loop to suppress the DC offset. To produce a high-pass pole, the DC offset is measured at the output by a low-pass filter and deducted at the input. To alleviate the DC offset, the schematic shown in Fig.~\ref{neural_amplifier}(b) leverages an integrator with a large time constant as an active low-pass filter~\cite{RN471, gosselin2007low}. This architecture does not require a large capacitor to provide substantial mid-band gain, but suffers from process variations. Resistance variation and the main amplifier's open-loop gain also have an impact on the high-pass pole. Furthermore, the amplifier in the feedback loop results in more power dissipation.

\subsubsection{Neural Amplifier Design Techniques}
Various circuit techniques have been developed to enhance neural amplifier performance. Among them, the chopping technique is widely used to amplify low-frequency signals since it can diminish offset and 1/f noise~\cite{RN147, RN638}. Fig.~\ref{neural_amplifier}(d) shows the diagram of chopping technique, which is elaborated in Section IV about noise reduction techniques. Other techniques include impedance bootstrapping \cite{RN630} and current-reuse techniques \cite{RN196}. These techniques are used to either improve the input impedance to prevent signal attenuation at the electrode-tissue interface or minimize circuit noise and power consumption for better SNR and noise efficiency factor (NEF). Additionally, neural amplifiers are required to be multiplexed using TDM or FDM to reduce power consumption of multi-channel neural interfaces. Crosstalk noise of amplified multi-channel signals also needs to be solved by dedicated techniques.
\vspace{-0.5cm}

\subsection{Analog to Digital Converters}
\vspace{-0.1cm}
This subsection explains the widely-used successive approximation register (SAR) ADC, followed by an active research topic on direct digitalization converters. Continuous-time event-driven level-crossing ADC (LC-ADC) is also discussed to explore its potential in low-power and low-frequency neural signal digitalization. Additionally, logarithmic pipeline ADC \cite{RN604} and single-slope ADC \cite{RN217} have been utilized for neural recording. However, this tutorial brief does not elaborate on these architectures because they are not popular in this field.

\begin{figure}[t]
	\centering
	\includegraphics[width=3in]{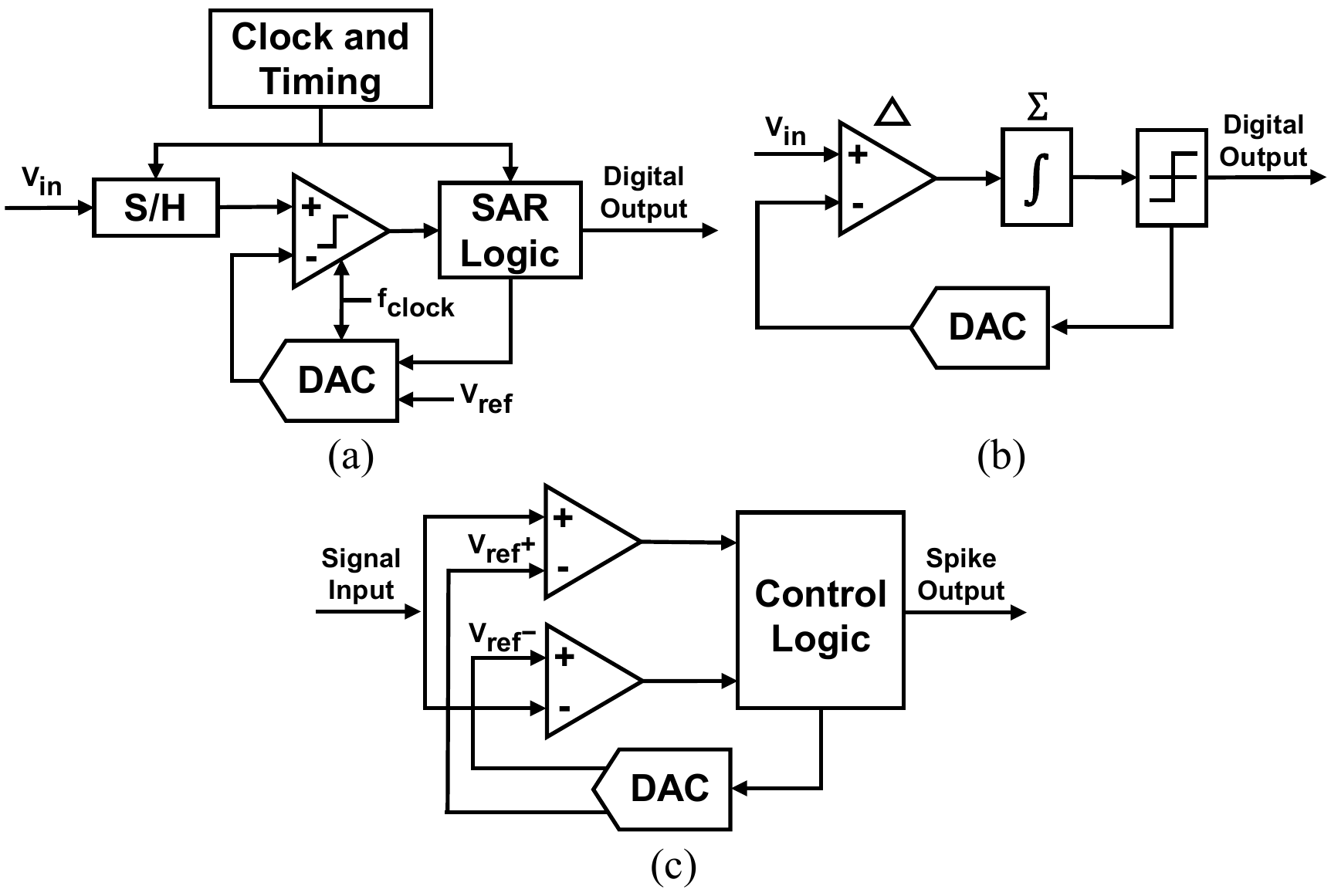}
	\vspace{-0.4cm}
	\caption{Basic structure of ADCs: (a) SAR ADC, (b) 1st order $\Delta$$\Sigma$ modulator, (c) LC-ADC.}
	\label{ADC}
	\vspace{-0.7cm}
\end{figure}

\subsubsection{SAR ADC}
The prevalent ADC used in neural recording is SAR ADC with moderate resolution of 8-10 bits, relatively low sampling speed of 1-500 kS/s, and typically 8-16 time-division multiplexing \cite{RN7}. Fig.~\ref{ADC}(a) shows the basic structure of SAR ADC. For a typical SAR ADC, one clock cycle is required to sample the input during the sampling phase, and one clock cycle is needed to determine each bit of the digital output in the conversion phase. As a result, digitizing an N-bit SAR ADC's input analog value typically takes N+1 clock cycles.

The major power sources of SAR ADC include capacitive DAC networks, comparators, and digital control circuits. Researchers have proposed energy-efficient switching methods to reduce DAC switching energy, such as monotonic capacitor switching \cite{liu201010} and adaptive-reset switching \cite{song20190}. Furthermore, monotonic capacitor switching SAR ADC \cite{liu201010} decreases the total capacitance by about 50\% compared to conventional SAR ADC. It is highly desirable in implant devices due to its low chip area consumption. Another direction to power reduction is to adjust the switching activity of ADC by considering characteristics of input bio-signals. The LSB-first SAR ADC is proposed in \cite{yaul201410} to reduce needed switching cycles with slowly varying input signals. Bypass switching techniques \cite{huang20121}\cite{ou2019energy} and dynamic tracking algorithms \cite{zhang202012} are also developed to skip redundant conversion steps and realize predictive digitization. Additionally, the scaling down of technology nodes enables faster devices and higher time-domain resolution, making time-domain comparators increasingly popular in low-power SAR ADCs. Voltage-controlled oscillator (VCO) based closed-loop \cite{ding20190} and voltage-controlled delay line (VCDL) based open-loop time-domain comparators \cite{zhou202112} are proposed and demonstrated to achieve high power efficiency.

\subsubsection{Direct Digitalization Converters}
Direct digitalization converters aim to digitalize raw neural signals directly without a low-noise amplification stage to reduce power and area overhead. Low-noise performance is achieved by oversampling $\Delta$$\Sigma$ ADC with intrinsic noise shaping capabilities. Furthermore, as technology node progresses, the scalability of $\Delta$$\Sigma$ ADCs allows for substantial chip area reduction, making $\Delta$$\Sigma$ ADC based direct digitization increasingly popular.

In 2016, Greenwald et al. developed a bidirectional neural interface with a 4-channel biopotential ADC (bioADC) implemented using a first-order $\Delta$$\Sigma$ modulator~\cite{RN55}. The microvolt neural signals are directly digitalized by a bioADC without a low-noise amplification stage. As shown in Fig.~\ref{ADC}(b), the input analog signal firstly goes through a $\Delta$-stage and is subtracted by the value from DAC. The following $\Sigma$-stage could be implemented by an operational transconductance amplifier (OTA) loaded with an output capacitor, such as $Gm-C$ integrators. A 1-bit quantizer connected with the $\Sigma$-stage generates digital output and controls the DAC output value. 

Direct digitalization converters offer efficient methods for removing electrode DC offset (EDO). In \cite{RN55}, the 1-bit output of the modulator is filtered by a discrete-time $\Delta$ integrator to extract the DC offset that is subtracted from the input by a DAC. However, this DC-servo loop based method suffers from limited input range and eliminates only $\pm$30 mV EDO. Another first-order $\Delta^{2}$$\Sigma$ modulator \cite{RN56} achieves rail-to-rail EDO removal with an integrator in the feedback path. The integrator's output is routed back into the input of the forward-path, allowing them only to digitize the voltage difference between two successive samples and remove EDO of the output thanks to signal shaping. A reset integrator is used to reconstruct the signal without EDO. Similarly, a second-order $\Delta$$\Sigma$-$\Sigma$ modulator \cite{RN575} is proposed to utilize an extra $\Sigma$-stage in the feedback path of the modulator to mitigate large EDO. Apart from the $\Delta$$\Sigma$ ADC, VCO based direct digitalization converters are also presented in \cite{RN607}\cite{liu2022hybrid}. A transconductor is typically utilized to transform the input voltage signal to current output for driving a ring oscillator. Large input signals are able to be sampled without saturating the VCO because only the phase is digitized by the ring oscillator. However, digital calibration techniques are usually required to address the intrinsic nonlinearity of the ring oscillator to improve the dynamic range and noise performance.

\subsubsection{LC-ADC}

\begin{figure}[t]
	\centering
	\includegraphics[width=3.5in]{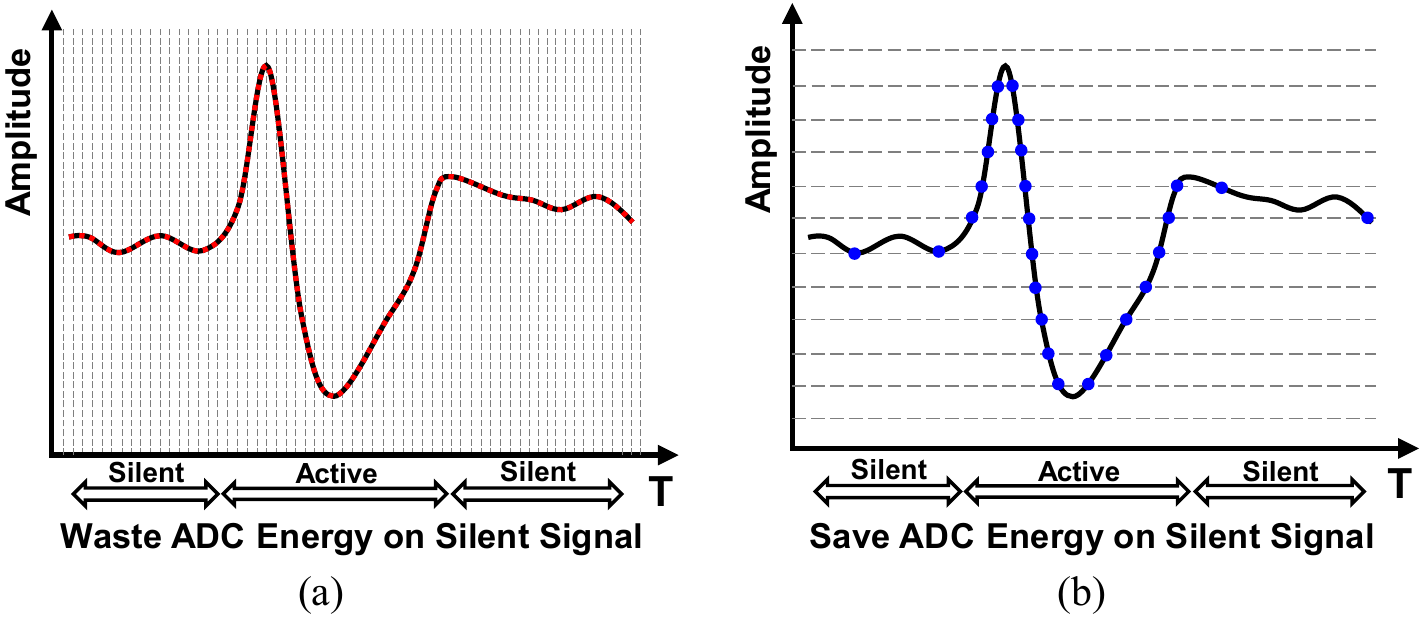}
	\vspace{-0.8cm}
	\caption{(a) Nyquist sampling; (b) Level-crossing sampling.}
	\label{LC_sampling}
	\vspace{-0.6cm}
\end{figure}

Neural signals are sparse with low-frequency and specific features. As presented in Fig.~\ref{LC_sampling}, neural spikes include silent period that is unimportant to feature extraction.  However, as shown in Fig.~\ref{LC_sampling}(a), typical uniform Nyquist sampling continuously samples neural spikes without considering their sparsity, squandering ADC energy during the silent interval. LC-ADC, on the contrary, is a nonuniform sampling frequency continuous-time event-driven data converter. It makes use of the sparsity of neural signals by sampling signals only when predetermined thresholds are passed. As described in Fig.~\ref{LC_sampling}(b), when the signal is silent and does not cross levels, there are no sampling points, which reduces the number of sampled data points and ADC energy consumption.

The basic LC-ADC architecture is shown in Fig.~\ref{ADC}(c). The input signal's voltage range is split into preset quantization levels denoted by $V_{ref+}$ and $V_{ref-}$. LSB is the numerical distance between two successive quantization levels, and is computed using the  following equation:
\vspace{-0.2cm}
\begin{equation}
LSB=\frac{A_{FS}}{2^{M}}
\label{equ:LCADC}
\vspace{-0.1cm}
\end{equation}
where $A_{FS}$ indicates the input signal's voltage amplitude range. The LC-ADC resolution bit is M, yielding ${2^{M}}$ quantization levels. When the quantization levels are up or down crossed, the LC-ADC samples the input signal and generates spike output.

Circuit-level LC-ADC development is an active research topic \cite{RN616, RN613, RN612, RN585, RN614}. Recently, LC-ADC has been applied to hardware-efficient neural spike sorting \cite{RN610} and real-time detection of high-frequency oscillations in intracranial EEG from epilepsy patients \cite{RN611}. 
\vspace{-0.4cm}

\subsection{Compressive Sensing}
Compressive sensing (CS) is actively employed in multi-channel neural recording systems to decrease power consumption and wireless data transfer rate by exploiting signal sparsity. Introduced in 2006, it facilitates sub-Nyquist sampling and near-lossless sparse signal reconstruction, enabling the acquisition of compressed data directly~\cite{RN620, RN619}. Most bio-signals are sparse in the Gabor, time, or wavelet domains \cite{RN7}, making them applicable to compressive sensing. 

The core principle of compressive sensing is a linear conversion of an N-dimension input signal (X) into an M-dimension measurement sequence (Y) using a measurement matrix $\Phi$ of a size M×N (Y = $\Phi$X). M is required to be smaller than N, reflecting data compression from N-dimension sample to M-dimension sequence. Because the equation is underdetermined, possible solutions for X are unlimited. Given a sparse X, the sparsest one has a high likelihood of being the right solution. As shown in \cite{RN619}, to execute signal reconstruction of sparse signal, a random measurement matrix $\Phi$ as a general encoder and considerable input samples X are required. The use of a pseudo-random Bernoulli matrix with every element $\Phi_{m,n}$ as $\pm$1 is a generic way to enable an improved circuit implementation of $\Phi$ \cite{RN621}. 

\begin{figure}[t]
	\centering
	\includegraphics[width=2.4in]{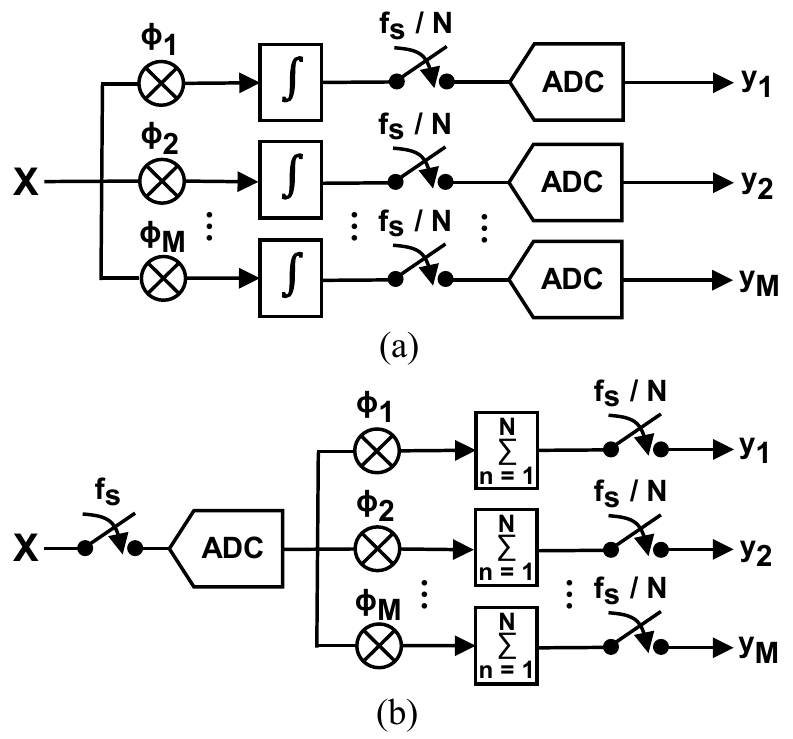}
	\vspace{-0.3cm}
	\caption{Block diagram of CS: (a) analog CS, (b) digital CS.}
	\label{CS}
	\vspace{-0.6cm}
\end{figure}

As illustrated in Fig.~\ref{CS}, both analog and digital domains are explored for the implementation of a CS encoder. Fig.~\ref{CS}(a) shows the block diagram of a CS encoder with analog implementation proposed in \cite{RN648, RN625, RN624}. The methods in \cite{RN624} leverage the spatial sparsity of intracranial EEG. \cite{RN625} uses the EEG signal's sparsity in the Gabor domain. Fig.~\ref{CS}(b) reveals the block diagram of digitally-implemented CS core. Digital implementation of CS encoder is explored in \cite{RN623, RN193}. Moreover, dedicated high-throughput and energy-efficient processors are proposed for signal reconstruction in \cite{RN626, RN627}. 
\vspace{-0.3cm}

\section{Noise Reduction Techniques}
First-stage low-noise amplifiers are crucial as they occupy the most significant noise contribution in neural recording. Flicker noise is dominant in low-frequency neural signal acquisition systems, because the noise level is inversely proportional to frequency. In low-noise amplifiers, discrete JFET devices are frequently employed in the input stages as preamplifiers because of their superior flicker noise characteristics \cite{RN642, RN641}. Nevertheless, JFETs are not typically available in standard CMOS processes \cite{RN643}. Researchers have used passive load \cite{RN473} and large-size transistors \cite{RN639} to alleviate the flicker noise effect. In this section, two popular noise reduction techniques, autozeroing (AZ) and chopper stabilization, are investigated.

\begin{figure}[t]
	\centering
	\includegraphics[width=3.5in]{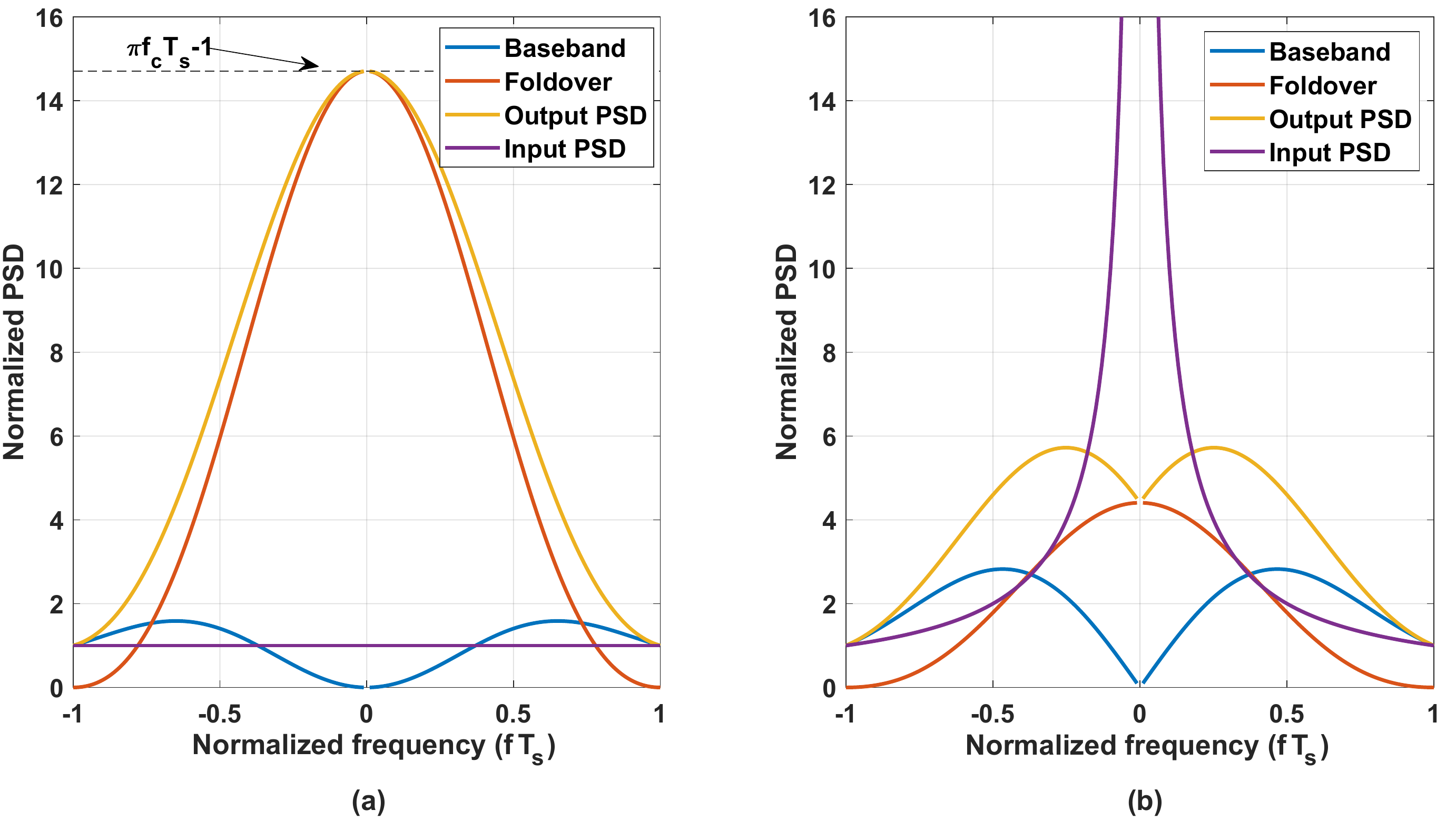}
	\vspace{-0.8cm}
	\caption{Analysis of autozeroing process effects on filtered (a) white noise, (b) flicker noise. The bandwidth of the filter is five times greater than the sampling frequency ($f_{c}$$T_{s}$=5). The corner frequency is the same as the sampling frequency ($f_{k}$$T_{s}$=1). (adapted from \cite{RN645}.)}
	\label{autozeroing}
	\vspace{-0.4cm}
\end{figure}

\vspace{-0.5cm}
\subsection{Autozeroing and Correlated Double Sampling}
A two-step process is involved in autozeroing technique: sampling and processing. Throughout the sampling phase, undesired signals including offset and noise are sampled and held. The held voltage is deducted from the contaminated signal during the processing phase. This method efficiently removes constant offsets and low-frequency noise. However, the efficiency of this method depends on the correlation between the sampled unwanted signal and the instantaneous unwanted signal present in the contaminated signal at the time of subtraction. High-frequency noise cannot be efficiently reduced due to the weak correlation between successive samples. 

Typically, the noise power spectrum density (PSD) of an autozero voltage can be separated into two parts: the baseband noise reduced by the AZ process and the fold-over noise caused by aliasing. Fig.~\ref{autozeroing} illustrates the effect of AZ on thermal and flicker noise. Both noise sources are suppressed in their baseband contents, but their high-frequency components are fold-over as a result of under-sampling. In most cases, the thermal noise fold-over component is dominant in the Nyquist band ($\left | fT_{s} \right |$$\leq$0.5). In this case, the amount of foldover component equals the initial thermal noise of the amplifier multiplied by the corresponding noise bandwidth to the Nyquist band ratio. Consequently, the production of this fold-over component causes a high noise floor~\cite{RN645}. Correlated double sampling (CDS) employs autozeroing followed by S/H. This technique is widely used in switched-capacitor (SC) circuits. Although a CDS circuit produces an output signal of S/H, its effect on amplifier noise is comparable to that of the AZ process.
\vspace{-0.6cm}

\begin{figure}[t]
	\centering
	\includegraphics[width=3.5in]{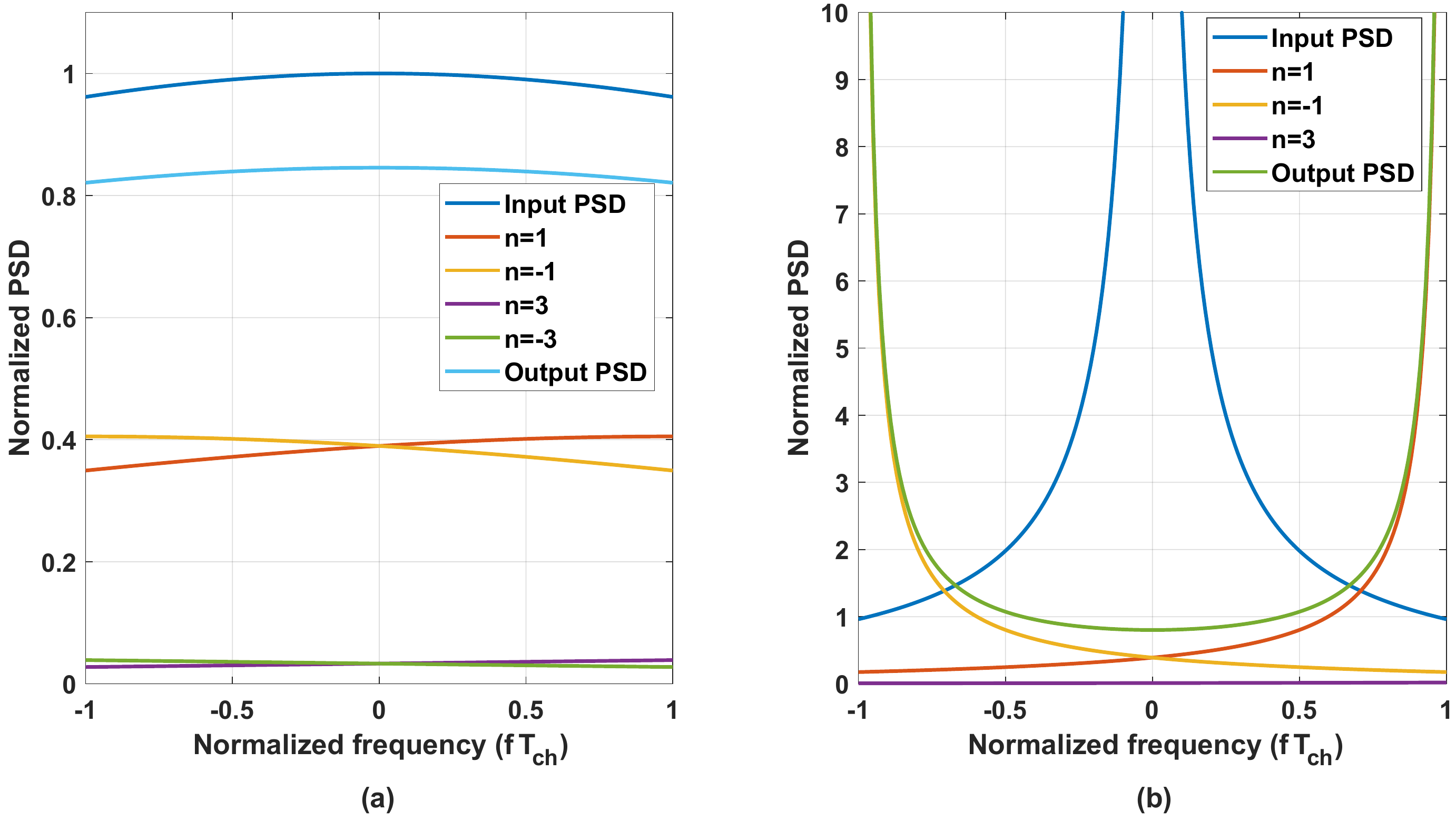}
	\vspace{-0.8cm}
	\caption{Analysis of the chopper stabilization technique effects on filtered (a) white noise, and (b) flicker noise. In this case, the bandwidth of the filter is five times greater than the chopper carrier frequency ($f_{c}$$T_{ch}$=5). Corner frequency is the same as chopper frequency ($f_{k}$$T_{ch}$=1). (adapted from \cite{RN645}.)}
	\label{chopper}
	\vspace{-0.6cm}
\end{figure}

\subsection{Chopper Stabilization Technique}
In the chopper stabilization process, two synchronized choppers with reversed polarity are utilized to realize signal modulation and demodulation. Each chopper is made up of four MOSFET switches. The input DC signal is modulated into a square wave by the input chopper, which is demodulated back to DC after amplification. The spectrum shows that the input signal is shifted to the even harmonics of the carrier frequency, whereas the noise generated by the amplifiers is transposed to the odd harmonics of the carrier frequency. The PSD of the output noise can be calculated as
\vspace{-0.2cm}
\begin{equation}
S_{C S}(f)=\left(\frac{2}{\pi}\right)^{2} \sum_{n=-\infty \atop n~o d d}^{+\infty} \frac{1}{n^{2}} S_{i}\left(f-\frac{n}{T_{c h}}\right)
\label{equ:chopper1}
\vspace{-0.1cm}
\end{equation}
where $T_{ch}$ is the period of the carrier and $S_{i}$ is the PSD of the input noise. The thermal noise in the Nyquist band ($\left | fT_{ch} \right |$$\leq$0.5) can be approximated as
\vspace{-0.3cm}
\begin{equation}
S_{t_{CS}} \cong S_{0}\left(1-\frac{\tanh \left(\frac{\pi}{2} f_{c} T_{c h}\right)}{\frac{\pi}{2} f_{c} T_{c h}}\right)
\label{equ:chopper2}
\vspace{-0.2cm}
\end{equation}
where $S_{0}$ is the thermal noise power of the amplifier. According to Fig.~\ref{chopper}(a), the baseband PSD arising from thermal noise is almost constant and smaller than that originating from the initial thermal noise, asymptotically approaching it when the ratio of cutoff frequency to chopper frequency is large. The chopper modulation approach, unlike the autozeroing technique, does not cause aliases in the high-frequency noise. Because the noise is neither sampled nor held, it is reversed repeatedly without modifying the overall noise features in the time domain.

The impact of chopper modulation on flicker noise is presented in Fig.~\ref{chopper}(b), showing that the flicker noise pole vanishes from the baseband after being transposed to the chopper frequency and its odd harmonics. It has also been proven that a white noise constituent within the baseband could be used to estimate the PSD of chopped flicker noise as \cite{RN645}
\vspace{-0.2cm}
\begin{equation}
S_{f_{C S}} \cong \frac{8}{\pi^{2}} S_{0} f_{k} T_{ch}\sum_{n=1 \atop n~o d d}^{\infty} \frac{1}{n^{3}} \approx 0.8526 S_{0} f_{k} T_{ch}
\label{equ:chopper3}
\vspace{-0.2cm}
\end{equation}

For a normal amplifier input, the overall residual noise in the baseband can be calculated as
\vspace{-0.2cm}
\begin{equation}
S_{C S}(f) \cong S_{0}\left(1+0.8526 f_{k} T_{ch}\right)
\label{equ:chopper4}
\vspace{-0.2cm}
\end{equation}

When setting the chopper frequency the same as the amplifier corner frequency, a good compromise is obtained. In this case, the white noise PSD increases by approximately 6 dB~\cite{RN645}. This technique is widely adopted in neural recording applications \cite{RN644, RN646}.
\vspace{-0.4cm}

\section{Future Trends and Challenges}
\subsection{Interaction With More Recording Channels}
By collecting neural recordings from different brain regions, it is possible to decode cognitive, motor, and sensory actions. Researchers have been actively investigating high spatiotemporal resolution recording systems with a significant increase in recording channels, effectively allowing brain-wide neural recording. Recently, recording systems like Neuropixels and Neuropixels 2.0 have been able to record 10000+ electrophysiological signals in parallel \cite{RN537, RN577}. Miniaturized high-density electrodes and  robust chronic implantable chips with intelligent algorithms are needed to make brain-wide neural recording more practical. Moreover, dedicated high-speed and low-power wireless communication is expected to pave the way for wireless wearable and implantable BMI.
\vspace{-0.5cm}

\subsection{Multimodal Neural Recording}
To explore diverse biomedical applications, it is essential to record multimodal neural signals. Traditional AFE is built for a certain type of recording signal without reconfigurability. Recently, AFEs with programmable gain, bandwidth, sampling rate, and reconfigurable modality have been developed for recording various neural signals on a single chip \cite{RN541, RN542}. Further innovative solutions for enhanced hardware accessibility and scaling ability are to be advocated in the future to address the design issues of ultra-low power consumption and compact chip area. Additionally, novel multiplexing techniques, such as code-division multiplexing described in \cite{RN481, RN578, RN152}, are expected to be investigated for future multi-channel wearable and implantable systems.
\vspace{-0.5cm}

\subsection{Analog-to-Information Conversion}
Conventional AFE encounters performance bottleneck in future ultra-high-density neural recording systems, as it wastes ADC energy on unnecessary data digitalization without utilizing the sparsity and features of neural signals.  To overcome this limitation, \cite{RN236} proposes the idea of analog-to-information, enabling the direct conversion of analog signals to decision-relevant information.  It not only minimizes the ADC conversion energy but also might significantly reduce power dissipation of wireless data transmission modules in wireless neural recording systems. Novel methods have been proposed, such as ADC integrating with feature extraction function \cite{RN286} and nonlinear signal-specific ADC \cite{RN649}. Moreover, the recent work \cite{RN230} emphasizes a viable strategy to explore event-driven level-crossing sampling techniques with intrinsic compression to achieve power-efficient analog-to-information systems.
\vspace{-0.4cm}

\section{Conclusion}
In this tutorial brief, the current status of neural recording is presented with emphasis on both system-level and design-level developments. At system-level, specifications of AFE and popular multiplexing techniques are discussed, highlighting time-division multiplexing and frequency-division multiplexing. In terms of circuit-level implementation, neural recording amplifiers and dedicated ADCs are explained, emphasizing direct digitalization converters and continuous-time event-driven LC-ADC. Compressive sensing and noise reduction techniques are also described. Moreover, future trends and challenges of neural recording are investigated, including interaction with more recording channels, multimodal neural recording, and analog-to-information conversion.

\section*{Acknowledgments}
This work is funded by Zhejiang Key R \& D Program project No. 2021C03002, and Zhejiang Leading Innovative and Entrepreneur Team Introduction Program No. 2020R01005.



\end{document}